# Nonextensive thermodynamic relations

Sumiyoshi Abe[a], S. Martínez[b, c], F. Pennini[b, c] and A. Plastino[b, c]

[a] *College of Science and Technology, Nihon University,*

*Funabashi, Chiba 274-8501, Japan*

[b] *Instituto de Física de La Plata, National University La Plata,*

*C.C. 727, 1900 La Plata, Argentina*

[c] *Argentine National Research Council (CONICET)*

The generalized zeroth law of thermodynamics indicates that the physical temperature in nonextensive statistical mechanics is different from the inverse of the Lagrange multiplier, $\beta$. This fact leads to modifications of some of thermodynamic relations for nonextensive systems. Here, taking the first law of thermodynamics and the Legendre transform structure as the basic premises, it is found that Clausius' definition of the thermodynamic entropy has to be appropriately modified, and accordingly the thermodynamic relations proposed by Tsallis, Mendes and Plastino [Physica A 261 (1998) 534] are also to be rectified. It is shown that the definition of specific heat and the equation of state remain form invariant. As an application, the classical gas model is reexamined and, in marked contrast with the previous result obtained by Abe [Phys. Lett. A 263 (1999) 424: Erratum A 267 (2000) 456] using the unphysical temperature and the unphysical pressure, the specific heat and the equation of state are found to be similar to those in ordinary extensive thermodynamics.

*PACS:* 05.70.-a; 05.20.-y; 05.90.+m

# 1. Introduction

It is generally believed that the framework of thermodynamics will remain unchanged forever even if the microscopic dynamical laws may be modified in the future. This optimism for robustness of thermodynamics might have its origin in the historical fact that statistical mechanics has been formulated in accordance with thermodynamics. However, if some of basic premises are removed from ordinary Boltzmann-Gibbs statistical mechanics, there is actually no *a priori* reason any more to be able to expect that all thermodynamic relations still remain unchanged. In such a situation, it is essential to specify a set of physical principles that should be preserved through this generalization.

In this paper, we discuss how removal of the assumption of extensivity of entropy from statistical mechanics leads to modifications of the thermodynamic relations. We develop this discussion by taking Tsallis' nonextensive statistical mechanics as an example [1-3]. A point of crucial importance is that, if entropy is nonextensive, the physical temperature is not simply the inverse of the Lagrange multiplier associated with the energy constraint but a variable correctly defined through the generalized zeroth law of thermodynamics [4-8]. The definition of the physical pressure also becomes different from the ordinary one. Taking into account these facts as well as the first law of thermodynamics and the Legendre transform structure [2,9,10], we show that Clausius' definition of the thermodynamic entropy has to be appropriately modified. We also show that the specific heat and the equation of state defined in terms of the physical temperature and the physical pressure remain form invariant. As an application, we reexamine the classical gas model. In marked contrast with the previous result using the "unphysical temperature" and the "unphysical pressure" [11], the specific heat and the equation of state are found to have the same forms as the ordinary ones in extensive thermodynamics.

## 2. Generalized zeroth law of thermodynamics

The concepts of temperature and pressure become nontrivial when entropy appears to be nonextensive. In such a case, it is necessary to reflect over the original ideas of macroscopic thermal and mechanical equilibria. In what follows, we consider this problem by taking the Tsallis nonextensive entropy as an example.

Suppose the total system be composed of two independent subsystems, $A$ and $B$, in thermal contact with each other. Then, the total Tsallis entropy satisfies [1]

$$S_q(A, B) = S_q(A) + S_q(B) + \frac{1-q}{k_T} S_q(A) S_q(B). \tag{1}$$

Here, $q$ is the positive entropic index whose deviation from unity measures the degree of nonextensivity. $k_T$ is a constant, which may depend on $q$, in general, and becomes the Boltzmann constant $k_B \cong 1.380658 \times 10^{-23} \, (J/K)$ in the extensive limit $q \to 1$. More generically, there is room for introducing system size dependence to $k_T$ [12], but in the present work $k_T$ is assumed to be system independent for the sake of simplicity.

Thermal equilibrium is characterized by the maximum total entropy state with the fixed total internal energy $U_q(A, B) = U_q(A) + U_q(B)$ [4-8]:

$$0 = \delta S_q(A, B) = \left[1 + \frac{1-q}{k_T} S_q(B)\right] \frac{\partial S_q(A)}{\partial U_q(A)} \delta U_q(A)$$
$$+ \left[1 + \frac{1-q}{k_T} S_q(A)\right] \frac{\partial S_q(B)}{\partial U_q(B)} \delta U_q(B). \qquad (2)$$

Since the total energy is fixed, we have the following separation of variables for the subsystems:

$$\frac{k_T \beta(A)}{1 + \frac{1-q}{k_T} S_q(A)} = \frac{k_T \beta(B)}{1 + \frac{1-q}{k_T} S_q(B)} \equiv k_T \beta^*, \qquad (3)$$

where

$$k_T \beta = \frac{\partial S_q}{\partial U_q}, \qquad (4)$$

which is actually the Lagrange multiplier associated with the energy constraint in Tsallis' nonextensive statistical mechanics [2]. Equation (3) defines the condition of thermal equilibrium as the equivalence relation for the generalized zeroth law of thermodynamics of nonextensive systems. It shows that, in contrast with the extensive case $(q \to 1)$, the physical temperature is not $(k_T \beta)^{-1}$ but

$$T_{phys} = \frac{1}{k_T \beta^*} = \left(1 + \frac{1-q}{k_T} S_q\right) \frac{1}{k_T \beta}. \qquad (5)$$

Similarly, we can define the physical pressure, $P_{phys}$, by further considering mechanical equilibrium [4]. In this case, the total entropy is maximized with fixing the total volume $V(A, B) = V(A) + V(B)$. The result is

$$\frac{\frac{\partial S_q(A)}{\partial V(A)}}{1 + \frac{1-q}{k_T} S_q(A)} = \frac{\frac{\partial S_q(B)}{\partial V(B)}}{1 + \frac{1-q}{k_T} S_q(B)} \equiv \frac{P_{phys}}{T_{phys}}. \qquad (6)$$

Therefore, we define the physical pressure as follows:

$$P_{\text{phys}} = \frac{T_{phys}}{1 + \frac{1-q}{k_T} S_q} \frac{\partial S_q}{\partial V}. \tag{7}$$

We shall see that the physical temperature and the physical pressure defined above necessarily lead to modification of Clausius' definition of the thermodynamic entropy.

3. **Nonextensive thermodynamic relations**

To establish the nonextensive thermodynamic relations, the principles we employ here are the thermodynamic Legendre transform structure and the first law of thermodynamics. The Legendre transform structure discussed so far in the nonextensive context offers the following expression for the generalized free energy [2]:

$$F'_q = U_q - \frac{1}{k_T \beta} S_q. \tag{8}$$

Note that the variable in front of the entropy is the inverse of the Lagrange multiplier. Accordingly, $F'_q$ is a function of the unphysical temperature. This is an unsatisfactory point, since all thermodynamic quantities should be expressed in terms of physical variables. Therefore, here we propose the following definition of the generalized free

energy:

$$F_q = U_q - T_{\text{phys}} \frac{k_T}{1-q} \ln\left(1 + \frac{1-q}{k_T} S_q\right). \tag{9}$$

With this definition, $F_q$ is seen to be in fact a function of $T_{\text{phys}}$. The derivative of $F_q$ yields

$$dF_q = dU_q - \frac{k_T}{1-q} \ln\left(1 + \frac{1-q}{k_T} S_q\right) dT_{\text{phys}} - T_{\text{phys}} \frac{dS_q}{1 + \frac{1-q}{k_T} S_q}. \tag{10}$$

Then, using the first law of thermodynamics

$$d'Q_q = dU_q + P_{\text{phys}} dV, \tag{11}$$

with $Q_q$ the quantity of heat, we find

$$dS_q = \left(1 + \frac{1-q}{k_T} S_q\right) \frac{d'Q_q}{T_{\text{phys}}}, \tag{12}$$

or equivalently,

$$\frac{k_T}{1-q} d \ln\left(1 + \frac{1-q}{k_T} S_q\right) = \frac{d'Q_q}{T_{\text{phys}}}. \qquad (13)$$

This shows how Clausius' definition of the thermodynamic entropy has to be modified for nonextensive systems.

From eqs. (10)-(12), we also have

$$\left(\frac{\partial F_q}{\partial T_{\text{phys}}}\right)_V = -\frac{k_T}{1-q} \ln\left(1 + \frac{1-q}{k_T} S_q\right) \qquad (14)$$

as well as the equation of state

$$P_{\text{phys}} = -\left(\frac{\partial F_q}{\partial V}\right)_{T_{\text{phys}}}. \qquad (15)$$

Note that equation (15) is actually consistent with eq. (7).

Substitution of eq. (14) into eq. (9) leads to

$$U_q = F_q - T_{\text{phys}} \left(\frac{\partial F_q}{\partial T_{\text{phys}}}\right)_V. \qquad (16)$$

Therefore, the specific heat is calculated to be

$$C_{qV} = \left(\frac{\partial U_q}{\partial T_{\text{phys}}}\right)_V = -T_{\text{phys}}\left(\frac{\partial^2 F_q}{\partial T_{\text{phys}}^2}\right)_V. \tag{17}$$

Thus, from eqs. (15) and (17), we see that both the equation of state and the specific heat remain form invariant, in view of the corresponding expressions in ordinary extensive thermodynamics.

## 4. Classical gas model revisited

The classical gas model in nonextensive statistical mechanics defines an unperturbed state of a system with a long-range interaction which may be treated perturbatively. This model has been solved analytically in Ref. [11]. There, the specific heat and the equation of state have been calculated using the unphysical temperature and the unphysical pressure, since the generalized zeroth law of thermodynamics for nonextensive statistical mechanics was not known at that time. An anomalous property was observed: the system has a negative specific heat.

Here, we reexamine the classical gas model based on the nonextensive thermodynamic relations derived in the preceding section.

The system Hamiltonian reads $H = \sum_{i=1}^{N} \mathbf{p}_i^2/2m$, where $\mathbf{p}_i$, $m$ and $N$ are the $D$-dimensional momentum of the $i$th particle, the common mass and the total number of the particles, respectively. The maximum Tsallis entropy distribution of this system is given by [5, 11]

$$f(\mathbf{p}_1, \mathbf{p}_2, \mathrm{L}, \mathbf{p}_N) = \frac{1}{Z_q(\beta)}\left[1-(1-q)(\beta/c)\left(\sum_{i=1}^{N}\frac{\mathbf{p}_i^2}{2m}-U_q\right)\right]^{1/(1-q)}, \qquad (18)$$

where the generalized partition function $Z_q(\beta)$, the internal energy $U_q$ and the factor $c$ are given by

$$Z_q(\beta) = \frac{V^N}{N! h^{DN}}\int \prod_{i=1}^{N} d^D\mathbf{p}_i \left[1-(1-q)(\beta/c)\left(\sum_{j=1}^{N}\frac{\mathbf{p}_j^2}{2m}-U_q\right)\right]^{1/(1-q)}, \qquad (19)$$

$$U_q = \frac{V^N}{c N! h^{DN}}\int \prod_{i=1}^{N} d^D\mathbf{p}_i \sum_{j=1}^{N}\frac{\mathbf{p}_j^2}{2m} f^q(\mathbf{p}_1, \mathbf{p}_2, \mathrm{L}, \mathbf{p}_N), \qquad (20)$$

$$c = \frac{V^N}{N! h^{DN}}\int \prod_{i=1}^{N} d^D\mathbf{p}_i\, f^q(\mathbf{p}_1, \mathbf{p}_2, \mathrm{L}, \mathbf{p}_N), \qquad (21)$$

respectively. In these equations, $h$ stands for the linear dimension of the elementary cell in phase space, and $\beta = (k_T)^{-1} \partial S_q / \partial U_q$ is the Lagrange multiplier associated with the energy constraint. The normalization condition

$$\frac{V^N}{N!h^{DN}} \int \prod_{i=1}^{N} d^D\mathbf{p}_i \, f(\mathbf{p}_1, \mathbf{p}_2, \mathrm{L}, \mathbf{p}_N) = 1 \qquad (22)$$

gives rise to the identical relation

$$c = \left[Z_q(\beta)\right]^{1-q}. \qquad (23)$$

It can be shown that this model is well defined in the two ranges of the entropic index: $0 < q < 1$ and $1 < q < 1 + 2/(DN)$. The latter is essentially the extensive limit ($q \to 1+0$) since $N$ is large. Therefore, as in Ref. [11], here we consider only the range

$$0 < q < 1. \qquad (24)$$

In this range, $Z_q(\beta)$, $U_q$ and $c$ are explicitly calculated to be [5,11]

$$Z_q(\beta) = \frac{\Gamma\left(\frac{2-q}{1-q}\right)}{\Gamma\left(\frac{2-q}{1-q} + \frac{DN}{2}\right)} \frac{V^N}{N!h^{DN}} \left[\frac{2\pi mc}{(1-q)\beta}\right]^{DN/2}$$

$$\times \left[1 + (1-q)\frac{\beta U_q}{c}\right]^{1/(1-q)+DN/2}, \qquad (25)$$

$$U_q = \frac{\dfrac{DN}{2\beta}}{\left[Z_q(\beta)\right]^q} \frac{\Gamma\left(\dfrac{2-q}{1-q}\right)}{\Gamma\left(\dfrac{2-q}{1-q}+\dfrac{DN}{2}\right)} \frac{V^N}{N!h^{DN}} \left[\frac{2\pi mc}{(1-q)\beta}\right]^{DN/2}$$

$$\times \left[1+(1-q)\frac{\beta U_q}{c}\right]^{1/(1-q)+DN/2}, \qquad (26)$$

$$c = \frac{1}{\left[Z_q(\beta)\right]^q} \frac{\Gamma\left(\dfrac{1}{1-q}\right)}{\Gamma\left(\dfrac{1}{1-q}+\dfrac{DN}{2}\right)} \frac{V^N}{N!h^{DN}} \left[\frac{2\pi mc}{(1-q)\beta}\right]^{DN/2}$$

$$\times \left[1+(1-q)\frac{\beta U_q}{c}\right]^{q/(1-q)+DN/2}, \qquad (27)$$

where $\Gamma(z)$ is the Euler gamma function. Noting the relation [11]

$$c = 1 + \frac{1-q}{k_T} S_q, \qquad (28)$$

and using eq. (23), the generalized free energy in eq. (9) is seen to be expressed as follows:

$$F_q = U_q - k_T T_{\text{phys}} \ln Z_q. \qquad (29)$$

In Refs. [2,11], $(k_T \beta)^{-1}$ was used as the temperature. However, now we know from the discussion in Sec. 2 that the physical temperature is $\left(k_T \beta^*\right)^{-1}$ in eq. (5), which is, in the present case, given by

$$T_{\text{phys}} = \frac{c}{k_T \beta}. \tag{30}$$

Now, from eqs. (25)-(27), we find a remarkable relation

$$\frac{\beta U_q}{c} = \frac{DN}{2}, \tag{31}$$

which yields the exact solution of the model [5,11]. Using the physical temperature for eq. (31), we obtain

$$U_q = \frac{DN}{2} k_T T_{\text{phys}}. \tag{32}$$

Therefore, the specific heat defined in eq. (17) is

$$C_{qV} = \frac{DN}{2} k_T. \tag{33}$$

This result is in complete parallel with the familiar result in ordinary Boltzmann-Gibbs statistical mechanics and should be compared with that in Ref. [11], where the negative specific heat is obtained using the unphysical temperature $(k_T \beta)^{-1} = (\partial S_q / \partial U_q)^{-1}$.

Finally, we calculate the physical pressure in eq. (15). Using eq. (32), we rewrite the generalized free energy in eq. (29) as follows:

$$F_q = \frac{DN}{2} k_T T_{\text{phys}} - k_T T_{\text{phys}} \ln Z_q. \tag{34}$$

Then, the physical pressure is

$$P_{\text{phys}} = -\left(\frac{\partial F_q}{\partial V}\right)_{T_{\text{phys}}} = k_T T_{\text{phys}} \left(\frac{\partial \ln Z_q}{\partial V}\right)_{T_{\text{phys}}}. \tag{35}$$

From eqs. (25), (30) and (31), we have

$$\left(\frac{\partial \ln Z_q}{\partial V}\right)_{T_{\text{phys}}} = \frac{N}{V}. \tag{36}$$

Therefore, we find

$$P_{\text{phys}} V = N k_T T_{\text{phys}}. \tag{37}$$

Thus, we arrive at a highly nontrivial conclusion that both the specific heat and the equation of state of the classical gas in nonextensive statistical mechanics have the same forms as those in ordinary extensive statistical mechanics.

5. **Conclusion**

We have defined the physical temperature and the physical pressure of nonextensive systems based on the generalized zeroth law of thermodynamics. In particular, we have emphasized that the physical temperature is different from the inverse of the Lagrange multiplier associated with the energy constraint. Then, imposing the Legendre transform structure and the first law of thermodynamics, we have constructed the generalized free energy and have established the nonextensive thermodynamic relations. In this way, the thermodynamic relations proposed by Tsallis, Mendes and Plastino [2] have been rectified. We have shown how Clausius' definition of the thermodynamic entropy has to be modified. We have found that the definition of specific heat and the equation of state remain form invariant under nonextensive generalization of thermodynamics. We have also shown that both the specific heat and the equation of state of the classical gas in nonextensive statistical mechanics have the same forms as those in ordinary extensive statistical mechanics, in marked contrast with the previous treatment using the inverse of the Lagrange multiplier as the temperature.

**Acknowledgments**

We would like to thank E. K. Lenzi for useful discussions. S. A. was by the Grant-in-Aid for Scientific Research of Japan Society for the Promotion of Science. S. M., F. P. and A. P. were supported by the National Research Council of Argentina (CONICET). F. P. Acknowledges financial support from UNLP.